\documentclass[fleqn,10pt,floatfix]{wlscirep}
\usepackage{bm}
\usepackage{bbold}
\newcommand{\mb}[1]{ { \mbox{\boldmath{$#1$}}}  } 

\title{Polarization of the Majorana quasiparticles\\ in the Rashba chain}

\author[1,*]{Maciej M. Ma\'ska}
\author[2,**]{Tadeusz Doma\'nski}
\affil[1]{Institute of Physics, University of Silesia, 40-007 Katowice, Poland}
\affil[2]{Institute of Physics, M. Curie-Sk{\l}odowska University, 20--031 Lublin, Poland}

\affil[*]{maciek@phys.us.edu.pl}
\affil[**]{doman@kft.umcs.lublin.pl}

\begin{abstract}
We demonstrate that the selective equal–spin Andreev reflection (SESAR) 
spectroscopy can be used in STM experiments to distinguish 
the zero–energy Majorana 
quasiparticles from the ordinary fermionic states of the Rashba chain. 
Such technique, designed for probing the $p$--wave superconductivity, 
could be applied to the intersite pairing of equal--spin electrons in 
the chain of magnetic Fe atoms deposited on the superconducting Pb substrate. 
Our calculations of the effective pairing amplitude for individual spin 
components imply the magnetically polarized Andreev conductance, which  
can be used to `filter' the Majorana quasiparticles from the ordinary 
in--gap states, although the pure spin current (i.e., perfect polarization) 
is impossible.
\end{abstract}
\begin{document}

\flushbottom
\maketitle
\thispagestyle{empty}

\section*{Introduction}

The topologically nontrivial superconducting state of one--dimensional (1D) chains 
\cite{Kitaev-2001} allows for a unique phenomenon of the {\em selective equal--spin 
Andreev reflection} (SESAR). This polarized Andreev spectroscopy has been proposed 
by J. J.\ He {\em et al} \cite{He-2014} as a useful tool for probing the Majorana states. 
SESAR measurements have indeed provided evidence for the zero--energy modes 
in vortices of the $p$--wave superconducting Bi$_{2}$Te$_{3}$/NbSe$_{2}$ 
heterostructures \cite{Sun-2016,Hu-2016}. Similar ideas have been also
considered for the Josephson--type junctions \cite{DasSarma-2015,Tanaka-2016} 
and ferromagnet--superconductor interfaces with the spin--orbit coupling 
\cite{Yuan-2017,Beiranvand-2016}. 
In this work we demonstrate that SESAR spectroscopy can test {\em inherent 
polarization} of the Majorana quasiparticles appearing at the edges of the 
Rashba chain. 
The parallel and perpendicular components of magnetically polarized Majorana states has 
initially been pointed out by D.\ Sticlet {\em et al}~\cite{Simon_2012} and their 
signatures have been recently studied by a number of authors
\cite{Kotetes-2015,Klinovaja-2017,Devillard-2017,YazdaniBernevig-2017}.
In this paper we show that magnetic polarization is detectable in STM experiments owing to SESAR processes, which in the subgap regime could distinguish 
the Majorana quasiparticles out the ordinary Shiba in-gap states.
We provide microscopic arguments explaining such polarization 
and confront our predictions with the experimental data obtained for Fe atom 
chain deposited on the surface of Pb superconductor by the STM technique 
with use of the magnetically polarized tip \cite{Yazdani-2017}.

%
%%%%%%%%%%%%%%%%%%%%%%%%%%%%%%%%%%%%%%%%%%%%%%%%%%
\begin{figure}[htb] % Fig 1
\centering
\centerline{\includegraphics[width=0.5\linewidth]{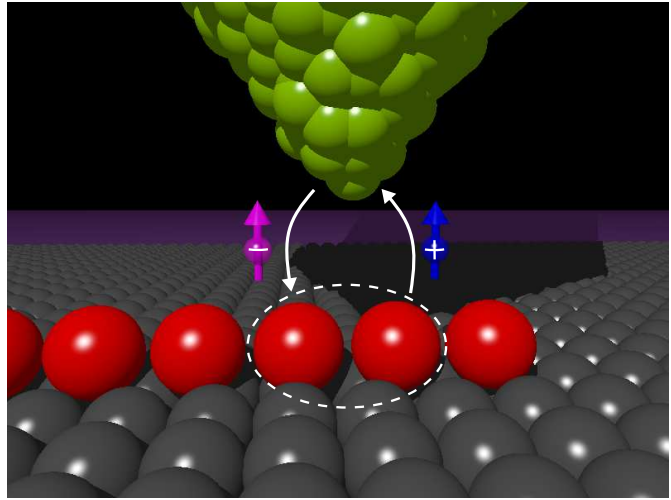}}
\caption{{\bf Schematic idea of SESAR.} This polarized Andreev spectroscopy 
can probe
the intersite pairing (represented by the dashed ellipse) of electrons 
on Fe atoms (red color) deposited on the $s$--wave bulk superconductor 
(gray) by using the magnetically polarized STM tip (green color).}
\label{fig1}
\end{figure}
%%%%%%%%%%%%%%%%%%%%%%%%%%%%%%%%%%%%%%%%%%%%%%%%%%
%

The underlying idea of SESAR for the aforementioned configuration is 
displayed in Fig.\ \ref{fig1}. This STM--type setup has been previously 
used by several experimental groups \cite{Yazdani-14,Kisiel-15,Franke-15},
however, ignoring the magnetic polarization. Recently A.\ Yazdani and 
coworkers \cite{Yazdani-2017} have measured the spin--resolved tunneling 
current and revealed substantial polarization of the zero--bias conductance 
in regions, where the Majorana quasiparticles exist. This fact can 
be interpreted within the popular microscopic model, taking into account
the Rashba and Zeeman interactions in  addition to the proximity--induced
pairing which can realistically capture a topography of the Majorana 
fermions \cite{Klinovaja-2016,DasSarma-2016,Stanescu-13,Simon_2012}. 
Using this model we have recently emphasized \cite{Maska-2016}, that 
amplitude of the intersite pairing (between identical spin electrons) differs 
several times for $\uparrow$ and $\downarrow$ sectors, respectively. Obviously, 
such effect should give rise to noticeable polarization of the Majorana quasiparticles 
near the chain edges. In practice, the low--energy features can be detected only
by the anomalous Andreev spectroscopy, as discussed in detail 
in Ref.\ \cite{Klinovaja-2016}. Since efficiency of the particle to hole 
conversion for the spin--polarized Andreev spectroscopy depends on 
the anomalous propagator $\langle\langle d_{i,\sigma};\, d_{i+1,\sigma}
\rangle\rangle_{\omega+i0^{+}}$, one should expect its non--vanishing value 
at $\omega=0$ nearby the chain edges. In  what follows we show, that this is 
really the case. We also argue, that SESAR could distinguish the Majorana 
from the ordinary fermionic quasiparticles.

\section*{Results}

\subsection*{Microscopic model} 
Nanoscopic chain of the magnetic Fe atoms deposited on the $s$--wave conventional
superconductor and probed by the  polarized STM tip (relevant 
to the experimental situation \cite{Yazdani-2017}) can be described by 
the Hamiltonian \cite{Klinovaja-2016,DasSarma-2016,Stanescu-13,Simon_2012}
% 
%\begin{eqnarray} 
$\hat{H} = \hat{H}_{\rm tip}  +  \hat{V}_{\rm tip-chain} + \hat{H}_{\rm chain} 
+  \hat{V}_{\rm chain-S} + \hat{H}_{\rm S}$.
%\label{model}
%\end{eqnarray} 
% 
We treat the STM tip $\hat{H}_{\rm tip}$ as a free fermion gas and focus on 
quasiparticle states of the atomic chain appearing deep inside the superconducting 
gap. Under such circumstances the superconducting reservoir would be responsible 
for the proximity induced on-site pairing  $\hat{H}_{\rm chain} +  
\hat{V}_{\rm chain-S} + \hat{H}_{\rm S} \rightarrow \hat{H}^{\rm prox}_{\rm chain}$
(for technical details see, e.g.,\ Appendix A in Ref.~\cite{Maska-2016}). 
%Hybridization between individual atoms 
%of the chain with STM tip is described $\hat{V}_{\rm tip-chain}$. 
In what follwos, we impose the constant couplings $\Gamma_{N}$ and 
$\Gamma_{S}$ to the STM tip and  superconducting substrate, 
respectively (see Fig.~\ref{fig1}).

The low--energy Hamiltonian is effectively given by 
 \cite{DasSarma-2016}
\begin{eqnarray}
\hat{H}^{\rm prox}_{\rm chain} &=& \sum_{i,j,\sigma}
(t_{ij} -\mu \delta_{i,j}) \hat{d}^{\dagger}_{i,\sigma} \hat{d}_{j,\sigma} 
+ \hat{H}_{\rm prox} +
 \hat{H}_{\rm Rashba}  +
\hat{H}_{\rm Zeeman} ,
% -\sum_{i} \left( \Delta_{i} \hat{d}^{\dagger}_{i\uparrow} % \hat{d}^{\dagger}_{i\downarrow} +\mbox{\rm h.c.}\right) ,
\label{chain_model}
\end{eqnarray}
where $\hat{d}_{i,\sigma}^{(\dag)}$ annihilates (creates) an electron 
with spin $\sigma$ at site $i$, $t_{ij}$ is the hopping integral and
$\mu$ is the chemical potential. The  proximity effect, responsible for 
the on--site (trivial) pairing, can be modeled as \cite{Stanescu-13}
\begin{eqnarray} 
\hat{H}_{\rm prox}= \Delta \left( \hat{d}_{i,\uparrow}^{\dagger} 
\hat{d}_{i,\downarrow}^{\dagger} + \hat{d}_{i,\downarrow}
\hat{d}_{i,\uparrow} \right) 
\label{large_Delta} 
\end{eqnarray} 
with the pairing potential $\Delta=\Gamma_{S}/2$.
In this scenario the intersite $p$--wave pairing 
is driven by the Rashba and the Zeeman interactions
\begin{eqnarray}
    \hat{H}_\mathrm{Rashba}&=&-\alpha\sum_{i,\sigma,\sigma'}\left[
    \hat{d}^{\dagger}_{i+1,\sigma}
    \left(i\sigma^y\right)_{\sigma\sigma'}\hat{d}_{i,\sigma'}+\mathrm{H.c.}\right],\\
    \hat{H}_\mathrm{Zeeman}&=&\frac{g\mu_\mathrm{B}B}{2}\sum_{i,\sigma,\sigma'}
    \hat{d}^{\dagger}_{i,\sigma}\left(\sigma^z\right)_{\sigma\sigma'}
    \hat{d}_{i,\sigma'} .
\end{eqnarray}
We assume the magnetic field to be aligned along $\hat{z}$--axis 
and impose the spin--orbit vector $\bm{\alpha}=(0,0,\alpha)$.

%
%%%%%%%%%%%%%%%%%%%%%%%%%%%%%%%%%%%%%%%%%%%%%%%%%%
\begin{figure}[htb] % Fig 2
\centering
\includegraphics[width=0.5\linewidth]{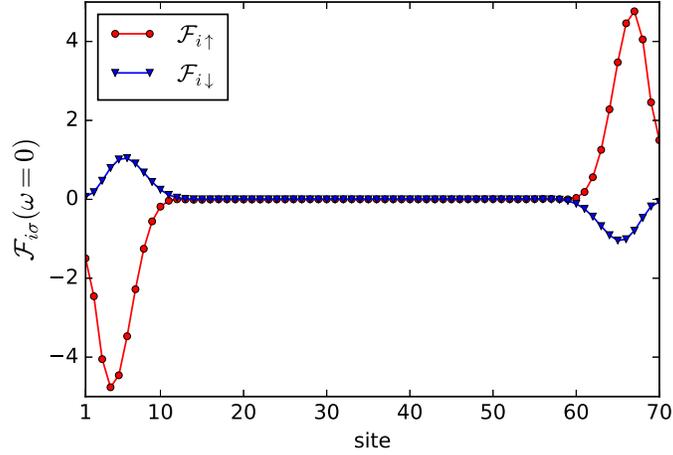}
\caption{{\bf Intrinsic polarization of Majorana quasiparticles.} 
The off--diagonal spectral function ${\cal F}_{i\sigma}
(\omega)$ obtained at zero energy ($\omega=0$) for the inter--site 
pairing of $\sigma$ spin electrons, using $\Delta=0.2t$, $\alpha=0.15t$, 
$\mu=-2.1t$, and $g\mu_{B}B/2=0.27t$.
%We have considered the Rashba chain comprising $L=70$ sites.
}
\label{G12}
\end{figure}
%%%%%%%%%%%%%%%%%%%%%%%%%%%%%%%%%%%%%%%%%%%%%%%%%%
%

\subsection*{Spin--polarized Majorana quasiparticles} 
In Fig. \ref{G12} we present spatial dependence of the off--diagonal
spectral function ${\cal F}_{i\sigma}(\omega)=-\frac{1}{\pi}
{\rm Im}\langle\langle \hat{d}_{i,\sigma}; \hat{d}_{i+1,\sigma} 
\rangle\rangle_{\omega+i0^{+}}$  obtained at zero energy for 
different spins $\uparrow$ and $\downarrow$, respectively. 
This anomalous spectral function is very instructive, because 
its sign exhibits {\em intrinsic polarization} of 
the Majorana modes (previously emphasized in Ref.\ \cite{Simon_2012})
whereas its absolute value can be probed by the SESAR spectroscopy 
(see the next paragraph). Concerning the magnitude, we clearly 
notice a quantitative difference (almost 5 times) between 
the spin $\uparrow$ and $\downarrow$ inter--site pairings. 
As regards the intrinsic polarization we observe that  
${\cal F}_{i\sigma}(\omega=0)$ changes its phase by $\pi$ 
between opposite sides of the Rashba chain and furthermore  
each of the spin sectors is characterized by opposite 
polarizations. This aspect resembles the results 
reported for the interface of ferromagnet/superconductor bilayers 
\cite{Levin-2016}. Such feature can be regarded as a hallmark of  the finite--size 
systems, because otherwise (i.e., in thermodynamic limit $L\rightarrow \infty$)
the off--diagonal spectral function would identically vanish at zero energy 
for both pairing channels. 

%
%%%%%%%%%%%%%%%%%%%%%%%%%%%%%%%%%%%%%%%%%%%%%%%%%%
\begin{figure}[htb] % Fig 3
\centering
\includegraphics[width=0.9\linewidth]{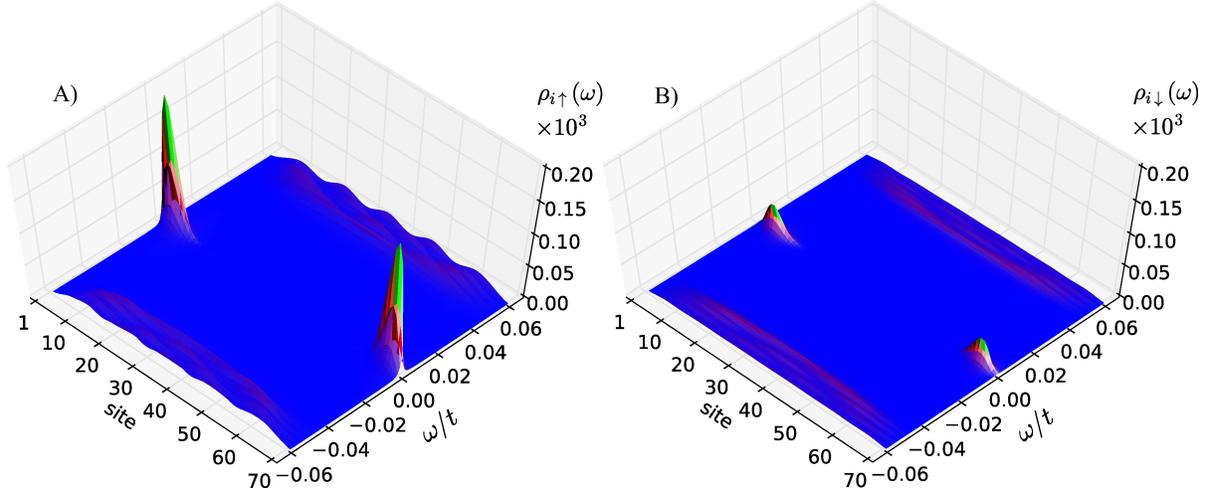}
\caption{{\bf Topography of the polarized quasiparticles.}
The spin--up (A) and spin--down (B) (diagonal) spectral functions $\rho_{i\sigma}
(\omega)$ determined at low energies which reveal, that the zero--energy 
(Majorana) quasiparticles are strongly polarized.}
\label{spin_spectr}
\end{figure}
%%%%%%%%%%%%%%%%%%%%%%%%%%%%%%%%%%%%%%%%%%%%%%%%%%
%

Fig.\ \ref{spin_spectr} illustrates the spatial profiles
of the spin--polarized (diagonal) spectral function 
$\rho_{i\sigma}(\omega)$.
% =-\frac{1}{\pi} {\rm Im}\langle\langle \hat{d}_{i,\sigma};
% \hat{d}^{\dagger}_{i,\sigma}  \rangle\rangle_{\omega+i0^{+}}$. 
As expected, we  notice
quantitative differences between the Majorana states appearing 
in $\uparrow$ and $\downarrow$ spin sectors, whereas their overall profiles   
seem to be pretty similar. Different magnitudes of these spin--polarized 
Majorana quasiparticles would show up in the SESAR measurements. 
% The pure spin current (discussed in Ref.~\onlinecite{He-2014}), however, 
% could not be realized because such quasiparticles are simultaneously 
% present for both the spin orientations.

\subsection*{The polarized Andreev transport}
By applying a bias voltage $V$ between the STM tip and the 
superconducting reservoir one would induce the nonequilibrium charge transport. 
Deep in a subgap regime (i.e.,\ for $|V| \ll \Delta/|e|$) such current 
is contributed solely by the Andreev scattering, when electrons from 
the STM tip are converted into the pairs, reflecting 
holes back to the STM tip. This process can be treated 
within the Landauer--B\"uttiker formalism.

We can express the nonmagnetic ($\gamma=0$) 
and magnetically polarized ($\gamma=\sigma$) Andreev currents by
the following  formula
\begin{eqnarray} 
I_{i}^{\gamma}(V) = \frac{e}{h} \int \!\!  d\omega \; T_{i}^{\gamma}(\omega)
\left[ f(\omega\!-\!eV)\!-\!f(\omega\!+\!eV)\right] ,
\label{I_A}
\end{eqnarray} 
where $f(x)=\left[1+\mbox{\rm exp}(x/k_{B}T)\right]$ stands for the 
Fermi--Dirac distribution function. These Andreev channels are 
characterized by various (dimensionless) transmittances, that 
can be  expressed via the local and non--local anomalous
Green's functions, respectively
\begin{eqnarray} 
T_{i}^{0}(\omega) &=& \Gamma_{N}^{2} \; \left( \left| \langle\langle 
\hat{d}_{i\uparrow};\, \hat{d}_{i\downarrow}\rangle\rangle \right|^{2}
+ \left| \langle\langle \hat{d}_{i\downarrow};\, \hat{d}_{i\uparrow}
\rangle\rangle \right|^{2} \right) ,
\label{conventional}\\
T_{i}^{\sigma}(\omega) &=&  \Gamma_{N}^{2} \; \left( \left| \langle\langle 
\hat{d}_{i\sigma};\, \hat{d}_{{i+1}\sigma}\rangle\rangle \right|^{2}
+ \left| \langle\langle \hat{d}_{i\sigma};\, \hat{d}_{{i-1}\sigma}
\rangle\rangle \right|^{2} \right) . 
\label{unconventional}
\end{eqnarray} 
Exceptionally, for the edge sites $i=1$ and $i=L$ the spin polarized transmittance is 
$T_{1}^{\sigma}(\omega) =  \Gamma_{N}^{2} \; \left| \langle\langle 
\hat{d}_{1\sigma};\, \hat{d}_{{2}\sigma}\rangle\rangle \right|^{2}$ and
$T_{L}^{\sigma}(\omega) =  \Gamma_{N}^{2} \; \left| \langle\langle 
\hat{d}_{L\sigma};\, \hat{d}_{{L-1}\sigma}\rangle\rangle \right|^{2}$. 
Derivation of formula (\ref{conventional}) is presented in section {\bf Methods}.
These off-diagonal Green's functions can be computed numerically from the 
Bogoliubov--de Gennes treatment of the Rashba chain (\ref{chain_model}).
Obviously, in experiments with the unpolarized STM tip 
\cite{Yazdani-14,Kisiel-15} the total current
contains all three components, i.e. $I_{i}(V)=\sum_{\gamma}I_{i}^{\gamma}(V)$.

%%%%%%%%%%%%%%%%%%%%%%%%%%%%%%%%%%%%%%%%%%%%%%%%%%
\begin{figure}[htb] % Fig 4
\centering
\includegraphics[width=\textwidth]{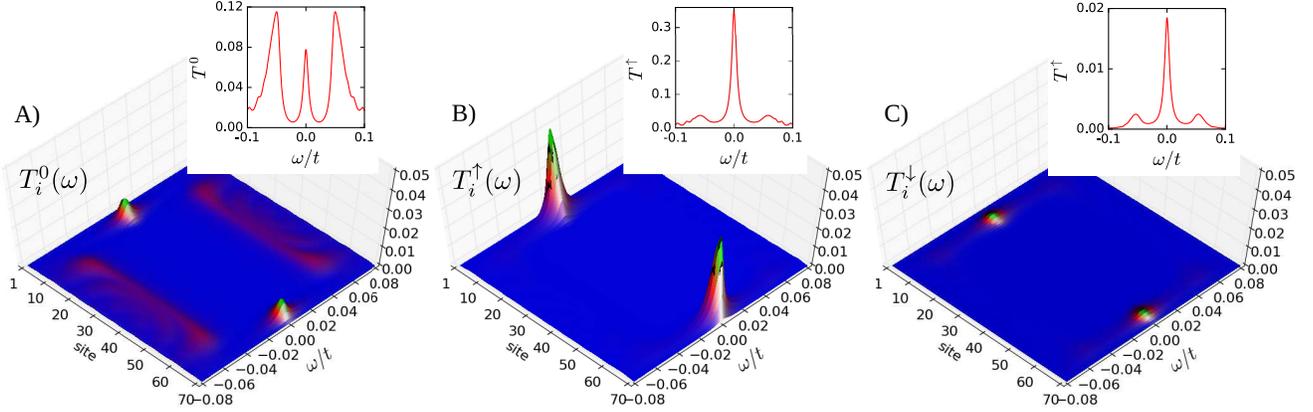}
\caption{{\bf Subgap transmittances.}
The spatially resolved transmittances $T_{i}^{\gamma}(\omega)$
obtained at low energies ($|\omega| \ll \Delta$) for the nonmagnetic 
$\gamma=0$ (panel A) and the spin--polarized Andreev reflections 
$\gamma=\uparrow$ (panel B) and $\gamma=\downarrow$ (panel C).
The insets display the transmittances summed 
over all lattice sites.}
\label{transmissions}
\end{figure}
%%%%%%%%%%%%%%%%%%%%%%%%%%%%%%%%%%%%%%%%%%%%%%%%%%

Figure \ref{transmissions} shows the energy--dependent transmittances $T_{i}^{\gamma}(\omega)$ 
obtained for the non--polarized ($\gamma=0$) and spin--polarized ($\gamma=\sigma$) 
Andreev channels. The difference between unpolarized and polarized 
transmittances is especially visible in the insets, where $T^{\gamma}(\omega)\equiv \sum_i T_{i}^{\gamma}(\omega)$ is plotted. In the case of $T^0(\omega)$ the  ordinary (finite-energy) Shiba states are are showing up (panel A), whereas in the polarized transmittances $T^{\uparrow,\downarrow}(\omega)$ the Majorana
quasiparticle plays the clearly dominat role (panels B and C).
%
%%%%%%%%%%%%%%%%%%%%%%%%%%%%%%%%%%%%%%%%%%%%%%%%%%
\begin{figure}[htb] % Fig 5
\centering
\includegraphics[width=\linewidth]{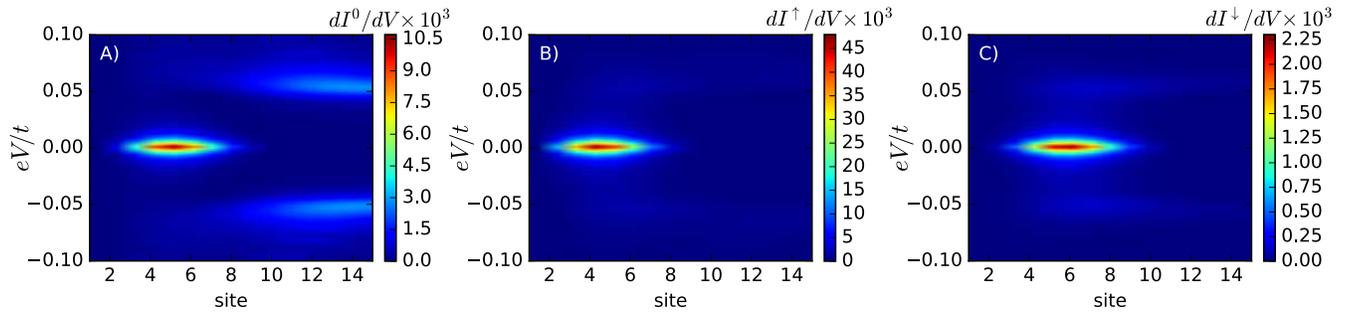}
\caption{{\bf Subgap conductances.}
False color plots of the differential conductance $dI^{\gamma}_{i}(V)/dV$ 
of the ordinary ($\gamma = 0$, panel A)  and the spin--resolved 
($\gamma=\uparrow$, panel B and $\gamma=\downarrow$, panel C) 
Andreev transport channels obtained at temperature 
$T=5\cdot 10^{-4}t$. The conductance is expressed in units $4e^{2}/h$. 
Plots B) and C) look very similar, but notice a strong difference 
in their scales.}
\label{conductances}
\end{figure}
%%%%%%%%%%%%%%%%%%%%%%%%%%%%%%%%%%%%%%%%%%%%%%%%%%

The corresponding conductances are presented in Fig.\ \ref{conductances}.
We notice that the differential conductance of the nonmagnetic Andreev reflections 
dominates well inside the Rashba chain at energies coinciding with the fermion Andreev/Shiba 
states. The SESAR, on the other hand, is efficient mainly near the Majorana modes whose spatial 
extent covers roughly 10 sites near the Rashba chain edges. In distinction to 
Ref.\ \cite{He-2014}, we observe that the spin--polarized currents are present 
for both spins ($\uparrow$ and $\downarrow$) but with significantly different magnitudes. 
Our results are relevant to the recent experimental 
data reported by  the Princeton group \cite{Yazdani-2017}. 
We have checked that the spin--polarized Majorana 
quasiparticles are robust upon varying the model parameters, although some additional subtle effects may be observed,
for instance the quantum oscillations \cite{Klinovaja-2016}. 

\begin{figure}[htb] % Fig 5a
\centering
\includegraphics[height=6cm]{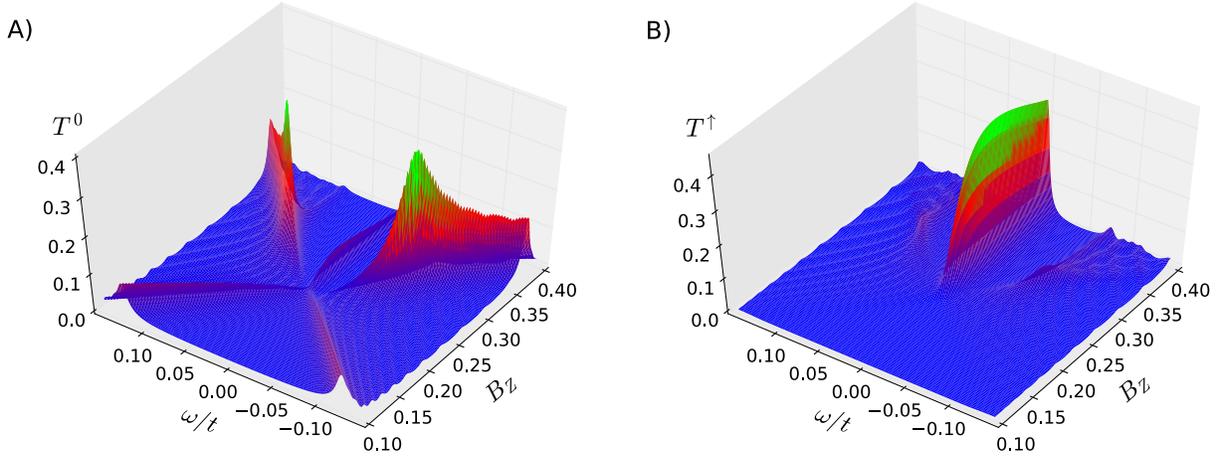}
\caption{{\bf Evolution of transmittances.} Unpolarized $T^0(\omega)$ (Panel A) and polarized $T^\uparrow(\omega)$ (Panel B) transmittances summed over all lattice sites as a function
of magnetic field. The topological phase starts around $B_Z=0.21$.}
\label{transmittances_notopo}
\end{figure}

The results presented in Fig.\ \ref{conductances} correspond to the topological regime. By varying the model parameters so that the system is driven to the topologically trivial phase, the zero-energy Majorana peak vanishes and the total transmittance in the spin-polarized channels is strongly suppressed. Such evolution from the topologically trivial to nontrivial state is presented in Fig.~\ref{transmittances_notopo}. 
Note that the polarized transmittance $T^\uparrow(\omega)$ vanishes almost completely outside the topological regime. In the topological regime the unpolarized transmittance of the Majorana peak is much smaller than the transmittance of the ordinary in-gap states that develop when the system enters the topological regime. On the other hand, the polarized transmittance of the Majorana peak is much larger than the ordinary in-gap states.

%
%%%%%%%%%%%%%%%%%%%%%%%%%%%%%%%%%%%%%%%%%%%%%%%%%%
%%%%%%%%%%%%%%%%%%%%%%%%%%%%%%%%%%%%%%%%%%%%%%%%%%

In summary, we emphasize that the net spin current
$I_{i}^{\rm spin}(V) = I_{i}^{\uparrow}(V)-I_{i}^{\downarrow}(V)$, 
attainable from the SESAR spectroscopy, is expected to acquire  
meaningful values of the spatially--resolved  conductance  
$G_{i}^{\rm spin}(V) = \partial I^{\rm spin}_{i} (V)/\partial V$  
only near the Majorana quasiparticles (what can be inferred by inspecting 
Fig.\ \ref{conductances}). SESAR can hence {\em filter} the 
Majorana from the ordinary Andreev/Shiba quasiparticles (which always exist 
in the Rashba chain). This  unique virtue of SESAR would be valuable 
for spotting the Majorana quasiparticles and investigating their topography.

\section*{Discussion}
We have studied the selective equal--spin Andreev spectroscopy (SESAR) 
which can empirically detect the polarized Majorana quasiparticles appearing 
at the edges of the Rashba chain. We have shown that different amplitudes of 
the inter--site equal--spin pairing imply the magnetic polarization of the Majorana 
states and yields the spin--dependent Andreev transport with substantially distinct 
probabilities in each spin components. Our theoretical results qualitatively 
agree with the recent finding by A.\ Yazdani \cite{Yazdani-2017}, who reported the 
spin--polarized features in the subgap spectroscopy.  Even though  
the pure spin current (discussed in Ref.\ \cite{He-2014}) is impossible --  the spin current conductance  $G_{i}^{\rm spin}(V)$ could nevertheless {\em filter} the Majorana 
quasiparticles from the ordinary Andreev/Shiba states. Our quantitative 
estimations clearly show also that the non--polarized and spin--polarized Andreev 
conductances are much smaller than the unitary limit value $2e^{2}/h$ as has 
been indeed observed by the STM \cite{Mourik-12,Yazdani-14,Kisiel-15} and 
by the tunneling measurements via heterojunctions \cite{Kouwenhoven-2016}. 

\section*{Methods}
Our calculations have been performed for the Rashba chain, comprising $L=70$ atoms. 
In most of the numerical calculations (except Figure \ref{transmittances_notopo}) we have used the following model parameters: magnitude of the induced pairing $\Delta=0.2\,t$, 
the spin--orbit coupling $\alpha=0.15\,t$, the chemical potential $\mu=-2.1\,t$, 
and the external magnetic field $g\mu_{B}B/2=0.27\,t$. Such a choice of parameters 
locates the system strictly in a topological regime \cite{Maska-2016}. 
The spin--resolved spectral functions, presented in Fig. \ref{spin_spectr},  have been calculated using the following definition
\begin{equation}
\rho_{i\sigma}(\omega)=-\frac{1}{\pi}
{\rm Im}\:\langle\langle \hat{d}_{i,\sigma}; \hat{d}^{\dagger}_{i,\sigma} 
\rangle\rangle_{\omega+i\Gamma_N/2},
\end{equation}
where $\Gamma_N$ is the coupling to the STM tip (assumed to be 
$\Gamma_{N}=0.01t$) and the Green function has been 
calculated numerically from 
$\hat{G}(\omega)=\left(\omega\,\mathbb{1}-\hat{H}^{\rm prox}_{\rm chain}\right)^{-1}$.
For $L$--site--long chain, the Hamiltonian $\hat{H}^{\rm prox}_{\rm chain}$ 
given by Eq.~(\ref{chain_model}), is $4L\times 4L$ complex matrix and
the currents in Eq.~(\ref{I_A}) have been calculated with a help of 8--point 
Gauss quadrature.

%%%from%%%%%%%%%%%%%%%%%%%%%%%%%%%%%%%%%%%%%%%%%%%%%%%%%%%%%%%%%%%%%%%%%%%%%%%%%%%
Let us outline a brief scheme for computing the charge tunneling current induced through 
$i$-th site of the chain coupled between the STM tip ($N$ electrode) 
and the superconducting substrate ($S$ electrode), for simplicity 
neglecting the inter-site hopping  $t_{ij}=0$. Using 
the Heisenberg equation we can express such current as
\begin{eqnarray}
I_{i}(V)= - e \frac{d}{dt} \left<  \hat{N}_{tip} \right>
= - e \left<  \frac{d}{dt} \hat{N}_{tip} \right>
=\frac{ie}{\hbar} \left< \left[ \hat{N}_{\rm tip}, \hat{V}_{{\rm tip}-i} \right] \right>,
\label{current_flow}
\end{eqnarray}
where $e$ stands for elementary charge,  $\hat{N}_{\rm tip}
=\sum_{\sigma,{\bf k}} \hat{c}^{\dagger}_{{\bf k},\sigma} 
\hat{c}_{{\bf k},\sigma}$ counts a number of electrons in STM tip,
and \\
$\hat{V}_{{\rm tip}-i} =\sum_{\sigma,{\bf k}} \left( V_{\bf k} 
\hat{d}^{\dagger}_{i,\sigma}\hat{c}_{{\bf k},\sigma} + \mbox{\rm h.c.}\right)$ denotes
the hybridization of $i$-th site with itinerant electrons of the tip. 
Since we are interested in the spin-resolved spectroscopy
let us exprees (\ref{current_flow}) as 
$I_{i}(V)=I_{i \uparrow}(V) + I_{i\downarrow}(V)$, where
\begin{eqnarray}
I_{i\sigma}(V)=  \frac{2e}{\hbar} \sum_{\bf k} 
\mbox{\rm Re} \left\{ V_{\bf k} \langle \langle
\hat{d}_{\sigma}(t); \hat{c}_{{\bf k}\sigma}^{\dagger}(t)
\rangle \rangle^{<} \right\} 
\label{current_form}
\end{eqnarray}
and the lesser Green's function is defined as 
$\langle \langle \hat{A}; \hat{B} \rangle \rangle^{<} \equiv
i\langle \hat{B} \hat{A} \rangle$. This mixed Green's function
can be determined using the Dyson equation
$\langle \langle \hat{A}; \hat{B} \rangle \rangle^{<}=
\langle \{ \hat{A}, \hat{B} \}_{+} \rangle +
\langle \langle \left[ \hat{A}, \hat{V} \right]; \hat{B} \rangle \rangle^{r}
g^{<}+\langle \langle \left[ \hat{A}, \hat{V} \right]; \hat{B} \rangle \rangle^{<}
g^{a}$. In our case, we obtain
\begin{eqnarray}
 \langle \langle \hat{d}_{\sigma}(t); \hat{c}_{{\bf k}\sigma}^{\dagger}(t)
\rangle \rangle^{<} = V_{\bf k}^{*} \int d\tau
\langle \langle
\hat{d}_{\sigma}(t); \hat{d}_{\sigma}^{\dagger}(\tau)
\rangle \rangle^{r} g_{\bf k}^{<} (t,\tau) +
V_{\bf k}^{*} \int d\tau
\langle \langle
\hat{d}_{\sigma}(t); \hat{d}_{\sigma}^{\dagger}(\tau)
\rangle \rangle^{<} g_{\bf k}^{a} (t,\tau)
\label{dyson}
\end{eqnarray}
with the bare Green's functions  $g_{\bf k}^{<} (t,\tau)=i \; 
f(\varepsilon_{\bf k}) e^{-i(\varepsilon_{\bf k} -eV)(t-\tau )}$ 
and $g_{\bf k}^{a} (t,\tau)=i \theta (-t+\tau) 
e^{-i(\varepsilon_{\bf k} -eV)(t-\tau )}$.

For studying the charge transfer in the low bias regime (comparable or 
smaller than energy gap $\Delta_{sc}$ of the superconducting electrode)   
we can impose constant couplings to the normal $\Gamma_{N} \equiv 2\pi 
\sum_{\bf k} |V_{\bf k}|^{2} \delta (\omega - \varepsilon_{\bf k})$
and superconducting electrode $\Gamma_{S} \equiv 2\pi 
\sum_{\bf q} |V_{\bf q}|^{2} \delta (\omega - \varepsilon_{\bf q})$.
Substituting (\ref{dyson}) to (\ref{current_form}) we get
\begin{eqnarray}
I_{i\sigma}(V)= - \; \frac{2e}{\hbar} \int \frac{d\omega}{2\pi}
\Gamma_{N}  \mbox{\rm Im} \left\{ \int_{-\infty}^{t} d \tau  
e^{i(\omega-eV)(t-\tau)} \left( \langle \langle
\hat{d}_{\sigma}(t); \hat{d}_{\sigma}^{\dagger}(\tau)
\rangle \rangle^{r} f(\omega) + \langle \langle
\hat{d}_{\sigma}(t); \hat{d}_{\sigma}^{\dagger}(\tau)
\rangle \rangle^{<} \right) \right\}
\label{current_long}
\end{eqnarray}
Introducing the Nambu notation  $\hat{\Psi}_{d}^{\dagger}
=(\hat{d}_{\uparrow}^{\dagger},\hat{d}_{\downarrow})$, $\hat{\Psi}_{d}=
(\hat{\Psi}_{d}^{\dagger})^{\dagger}$ we can define 
the matrix Green's function \\
${\mb G}_{d}(\tau,\tau')
\!=\!\langle\langle \hat{\Psi}_{d}(\tau); \hat{\Psi}_{d}^{\dagger}(\tau')
\rangle\rangle$ and recast expression (\ref{current_long}) as 
\begin{eqnarray}
I_{i\uparrow}(V)= - \; \frac{2e\Gamma_{N}}{h} \int d\omega\, 
\mbox{\rm Im} \left\{ \int_{-\infty}^{t} d \tau  
e^{i(\omega-eV)(t-\tau)} \left( {\mb G}_{d}(t,\tau )_{11}^{r}
 f(\omega) + {\mb G}_{d}(t,\tau )_{11}^{<} \right) \right\}
\label{current_short}
\end{eqnarray}
The lesser matrix Green's function obeys the  Keldysh
equation ${\mb G}^{<}=( {\mb 1} + {\mb G}^{r} {\mb \Sigma}^{r} )
{\mb g}^{<} ( {\mb 1} + {\mb G}^{a} {\mb \Sigma}^{a} ) +
{\mb G}^{r}{\mb \Sigma}^{<} {\mb G}^{a}$, where for brevity  
we dropped the temporal arguments. In our case the first term
vanishes, so we are left with
${\mb G}^{<}_{11} =  
{\mb G}^{r}_{11}  {\mb \Sigma}^{<}_{11}  {\mb G}^{a}_{11} +
{\mb G}^{r}_{11}  {\mb \Sigma}^{<}_{12}  {\mb G}^{a}_{21} +
{\mb G}^{r}_{12}  {\mb \Sigma}^{<}_{21}  {\mb G}^{a}_{11} +
{\mb G}^{r}_{12}  {\mb \Sigma}^{<}_{22}  {\mb G}^{a}_{21} 
$.
Using the explicit selfenergies ${\mb \Sigma}^{<}_{\alpha \beta}(t,\tau)$ 
we finally obtain the total current given by \cite{Sun-1999}
\begin{eqnarray}
I_{i}(V) = I^{0}_{i}(V) + I^{1}_{i}(V) ,
\label{general}
\end{eqnarray}
where the first contribution (Andreev  current)
\begin{eqnarray}
I^{0}_{i}(V) = \frac{e}{h} \int d\omega 
\; T^{0}_{i}(\omega) \;  \left[ f(\omega+eV) - f(\omega-eV) \right]
\label{And_curr}
\end{eqnarray}
describes processes, in which electrons from the normal STM tip
are scattered back to the same electrode holes, injecting Cooper 
pairs to the superconducting substrate. Its transmittance 
depends on the anomalous (off-diagonal) retarded Green's function 
\begin{eqnarray}
T^{0}_{i}(\omega)=\Gamma_{N}^{2} \left| 
\langle\langle \hat{d}_{i\uparrow} \hat{d}_{i\downarrow}\rangle
\rangle_{\omega}^{r} \right|^{2} 
\hspace{0.2cm} + \hspace{0.2cm} ' \uparrow  \hspace{0.1cm} 
\longleftrightarrow  \hspace{0.1cm} \downarrow ' .
\label{And_transm}
\end{eqnarray}
The other contribution appearing in equation (\ref{general}) takes the usual form
\begin{eqnarray}
I^{1}_{i}(V) = \frac{e}{\hbar} \int d\omega 
\; T^{1}_{i}(\omega) \; \left[ f(\omega+eV) - f(\omega) \right]
\end{eqnarray}
and its transmittance consists of three terms
\begin{eqnarray}
T^{1}_{i}(\omega)&=&\Gamma_{N}\Gamma_{S} \rho_{S}(\omega)
\left( \left| \langle
\langle \hat{d}_{i\uparrow} \hat{d}^{\dagger}_{i\uparrow}\rangle
\rangle_{\omega}^{r} \right|^{2} +\left| 
\langle\langle \hat{d}_{i\uparrow} \hat{d}_{i\downarrow}\rangle
\rangle_{\omega}^{r} \right|^{2} - \frac{2\Delta_{sc}}{|\omega|}
\mbox{\rm Re} \left[
\langle \langle \hat{d}_{i\uparrow} \hat{d}^{\dagger}_{i\uparrow}
\rangle \rangle_{\omega}^{r} \langle\langle \hat{d}_{i\uparrow} 
\hat{d}_{i\downarrow}\rangle \rangle_{\omega}^{r} \right] \right)
\hspace{0.2cm} + \hspace{0.2cm} ' \uparrow  \hspace{0.1cm} 
\longleftrightarrow  \hspace{0.1cm} \downarrow ' 
\end{eqnarray}
with $\rho_{S}(\omega)=\frac{|\omega|}{\sqrt{\omega^{2}-\Delta_{sc}^{2}}}
\theta(|\omega|\!-\!\Delta_{sc})$.
These terms correspond to the single particle tunneling,  
electron to hole conversion ("branch crossing" in the language 
of Blonder-Tinkham-Klapwijk approach) and electron to Copper 
pair scattering, respectively \cite{Sun-1999}. At zero temperature
$I^{1}_{i}(V)$ vanishes in the sub-gap regime  $e|V| < \Delta_{sc}$ 
for this reason the charge current can be transmitted solely 
via the Andreev channel.

Situation studied by us in the main text is a bit more complex, 
because of the inter-site $p$-wave pairing that activates the equal 
spin Andreev scattering processes. Their contribution to the subgap 
current can be expressed in the same way as (\ref{And_curr}) with 
straightforward generalization of the transmission (\ref{And_transm}).

%%%%%%%%%%%%%%%%%%%%%%%%%%%%%%%%%%%%%%%%%%%%%%%%%%%%%%%%%%%%%%%%%%%%%%%%%%%%%%%%

\section*{Acknowledgments}

We thank C.\ Bena, R. M.\ Lutchyn, J.\ Klinovaja, P.~Simon, and R.\ \v{Z}itko 
for discussions on the Majorana states and the Andreev spectroscopy. This work 
is supported by the National Science Centre (Poland) under the contracts 
DEC--2014/13/B/ST3/04451 (TD) and DEC--2013/11/B/ST3/00824 (MMM).

\section*{Author contributions statement}
T.D. posed the problem and prepared the first version of the manuscript. M.M.M. carried out the numerical calculations.
Both authors discussed the results and contributed to the final form of the paper.

\section*{Additional information}

\textbf{Competing financial interests} 
The authors declare no competing financial interests.

\end{document}